\documentclass[prd,amsmath,amssymb,floatfix,superscriptaddress,notitlepage,showpacs]{revtex4}
\usepackage{graphicx}
\usepackage{hyperref}
\usepackage{bm}
\usepackage{amssymb}

\begin{document}

\title{Generation of Coherent Structures After Cosmic Inflation}

\author{Marcelo Gleiser}
\email{mgleiser@dartmouth.edu}
\affiliation{Department of Physics and Astronomy, Dartmouth College,
Hanover, New Hampshire 03755, USA}

\author{Noah Graham}
\email{ngraham@middlebury.edu}
\affiliation{Department of Physics, Middlebury College,
Middlebury, Vermont 05753, USA}

\author{Nikitas Stamatopoulos}
\email{nstamato@dartmouth.edu}
\affiliation{Department of Physics and Astronomy, Dartmouth College,
Hanover, New Hampshire 03755, USA}

\date{\today}

\begin{abstract}
We investigate the nonlinear dynamics of hybrid inflation models, which are characterized by two real scalar fields interacting quadratically. We start by solving numerically the coupled Klein-Gordon equations in static Minkowski spacetime, searching for possible coherent structures. We find long-lived, localized configurations, which we identify as a new kind of oscillon. We demonstrate that these two-field oscillons allow for ``excited'' states with much longer lifetimes than those found in previous studies of single-field oscillons. We then solve the coupled field equations in an expanding Friedmann-Robertson-Walker spacetime, finding that as the field responsible for inflating the Universe rolls down to oscillate about its minimum, it triggers the formation of long-lived two-field oscillons, which can contribute up to 20\% of the total energy density of the Universe.  We show that these oscillons emerge for a wide range of parameters consistent with WMAP 7-year data. These objects contain total energy of about $25\times10^{20}$ GeV, localized in a region of approximate radius $6\times 10^{-26}$ cm. We argue that these structures could have played a key role during the reheating of the Universe.
\end{abstract}
\maketitle

\section{Introduction}
Inflation \cite{guth} is widely regarded as the mechanism behind the generation of the density fluctuations that eventually led to the formation of large scale structure in the Universe. In its simplest version, inflation is driven by a scalar field, the inflaton, which slowly rolls down its potential. While the total energy density is dominated by the inflaton's potential energy, the Universe grows exponentially fast. Eventually, the inflaton decays and reheats the Universe \cite{basset}, giving rise to a radiation-dominated era of expansion consistent with big bang nucleosynthesis \cite{bbn}. Given that we do not have a compelling candidate for the inflaton or for the fields it decays into during reheating, most of the work in inflationary cosmology focuses on the properties of different models and how they can be constrained by current data \cite{kolb,adshead}.

A particular class of slow-roll inflation models that is still viable by present constraints is provided by hybrid inflation \cite{linde_hybrid}, where in addition to the inflaton $\phi$, there is another field $\psi$, called the ``waterfall'' field. While the inflaton has a zero vev, the waterfall field has a nonzero vev. During inflation, the waterfall field is trapped at the minimum of its potential. After the inflaton crosses below a critical value $\phi_c$ during its slow roll, the curvature at the origin becomes negative and the waterfall field undergoes spontaneous symmetry breaking: both fields are quickly driven towards their vevs, effectively terminating inflation and setting the initial conditions for reheating \cite{bellido-linde}. 

The dynamics of reheating after hybrid inflation has been studied both analytically and numerically \cite{felder-preheating,copeland-preheating,basset}. In this paper, we report results of 3d numerical simulations of the coupled scalar field equations just after the end of hybrid inflation, as the waterfall field $\psi$ relaxes spinodally towards its vev and the inflaton starts to oscillate about its minimum. Because of the quadratic coupling between the two fields, nonlinear effects are prevalent. Indeed, inflaton ``hot spots'' were reported in \cite{copeland-preheating}, while ``condensate lumps'' were found for a class of supersymmetric models in \cite{mcdonald}. Here, we observe the emergence of long-lived, time-dependent field configurations, which we identify as a new class of two-field oscillons \cite{bogol,gleiser,copeland,fodor-honda}. If oscillons are sufficiently long-lived, that is, if their lifetime is at least of order of $H^{-1}$, they can have important consequences for the cosmological dynamics. The potential role of oscillons in cosmology has been investigated in 1d simulations \cite{graham_cos,farhi_cos,amin1d}, while 3d simulations have been carried out in the case of a single real field in a double-well potential \cite{3d_osc_temperature} and in the case of a $\phi^6$ potential \cite{amin3d} where the emergence of flat-top oscillons \cite{flat-top} was studied. We find that these two-field, ``hybrid oscillons'' are indeed very long-lived, and emerge for a wide range of model parameters. Given that the inflaton must couple to other fields, and that oscillons have been found in a wide variety of models, from Abelian-Higgs models \cite{gleiser-thor} to the standard model \cite{osc-gauge}, we should expect oscillonlike structures to emerge during preheating, considerably enriching the post-inflationary dynamics and the reheating of the Universe.

This paper is organized as follows. In the next section, we investigate the model with two coupled scalar fields, in the absence of cosmological expansion, and show that it admits spherically-symmetric oscillonlike solutions. These solutions are attractors in field configuration space: different initial conditions lead to long-lived, localized configurations with the same total energy. We also show that our model admits two different types of these oscillon solutions: the ``ground state,'' where oscillons have similar lifetimes and energies to the well-known single-field models \cite{gleiser,copeland,fodor-honda}, and a newly found ``excited state,'' where oscillon lifetimes can be three to four times as large as those for the ground state. We may think of the excited state as a metastable state with two possible decay channels: either it decays by directly radiating its energy to spatial infinity or, quite interestingly, it decays to the ground state configuration, where it remains for a while before radiating its energy to spatial infinity. This two-step decay inspires us to associate these higher-energy configurations with excited oscillon states. In Sec.~\ref{sec:model} we introduce the hybrid inflation model and constrain the parameters to be consistent with WMAP 7-year data \cite{WMAP}. In Sec.~\ref{sec:results} we report results from numerical simulations in an expanding Universe, calculating the fraction of energy in oscillons for a range of values of the ratio of coupling constants $g^2/\lambda$, where $g^2$ is the quadratic coupling between the two scalar fields and $\lambda$ is the quartic coupling of the waterfall field. Moreover, we show that for  $g^2/\lambda=2$, oscillon production is particularly efficient, with the fraction of energy in oscillons reaching approximately 20\%. We show that this effect is due to parametric resonance of a band of low $k$ modes of the waterfall field, arising from the resonant oscillation of its zero mode and of the inflaton about its minimum. We conclude with a summary of our results and directions for future work.    

\section{Oscillons in a Two Scalar Field Model}
\label{sec:1d}
In \cite{gleiser,copeland,fodor-honda}, spherically-symmetric oscillons were found in a single scalar field model with a symmetric (and also with asymmetric) double-well potential, starting from an initial Gaussian field profile that interpolated between the two minima of the potential. We perform a similar analysis in $(3+1)$ dimensions, now working with two scalar fields coupled quadratically, governed by a potential used in many hybrid inflation models

\begin{equation}
 V(\psi,\phi)=\frac{\lambda}{4}\left(\psi^2-\frac{M^2}{\lambda}\right)^2+\frac{1}{2}m^2\phi^2+\frac{1}{2}g^2\phi^2\psi^2.
\label{hybrid_potential}
\end{equation}
Defining dimensionless variables $\tilde{x}^{\mu}=Mx^{\mu}$, $\tilde{\psi}=\psi(M/\sqrt{\lambda})^{-1}$, $\tilde{\phi}=\phi(M/\sqrt{\lambda})^{-1}$, setting $\hbar=c=1$ and dropping the tildes, the equations of motion obeyed by the scalar fields become

\begin{equation}
 \ddot{\phi}-\nabla^2\phi = -\left(\frac{m^2}{M^2}\phi+\frac{g^2}{\lambda}\psi^2\phi\right)
\label{phi_eq_1d}
\end{equation}
\begin{equation}
 \ddot{\psi}-\nabla^2\psi = \left(\psi-\frac{g^2}{\lambda}\phi^2\psi-\psi^3\right),
\label{psi_eq_1d}
\end{equation}
where an overdot denotes the derivative with respect to dimensionless time. For the moment, we will only consider spherically-symmetric configurations such that $\nabla^2\rightarrow\partial^2/\partial r^2+(2/r)\partial/\partial r$ in Eqs.~\ref{phi_eq_1d} and ~\ref{psi_eq_1d}. We look for oscillon solutions starting from Gaussian initial profiles for both fields, $\psi(r,0)=2\exp(-r^2/R_{\psi}^2)-1$ and $\phi(r,0)=2\exp(-r^2/R_{\phi}^2)$ while setting $\dot{\phi}(r,0)=\dot{\psi}(r,0)=0$. The effective masses of the two fields (in units of $M$) are given by

\begin{equation}
 m_{\phi}^2=\frac{m^2}{M^2}+\frac{g^2}{\lambda}\psi^2
\end{equation}
\begin{equation}
 m_{\psi}^2=3\psi^2+\frac{g^2}{\lambda}\phi^2-1.
\label{field_masses}
\end{equation} 

We discretize the equations of motion using fourth-order spatial derivatives with lattice spacing $\Delta r=0.02$ and step forward in time using a fourth-order R\"{u}nge-Kutta method with $\Delta t=0.01$. We have verified that any further reduction of $\Delta r$ and $\Delta t$ does not affect our results. We add a damping term near the lattice boundary to absorb any radiation that may be reflected back and interfere with the oscillon evolution \cite{sornborger}. We ensure that no energy comes back into the oscillon region by computing the flow of energy through the ``energy shell'' and verifying that no energy goes back in.

The equations of motion contain two free parameters, as of yet unconstrained: the ratio of the square of the two mass scales $m^2/M^2$, and the ratio of couplings $g^2/\lambda$. Since the main goal of this paper is not to explore in detail the properties of two-field oscillons (a worthy task left for a future publication) but to show that they exist and explore their cosmological implications, in this section we fix the parameters as $m^2/M^2=1.5\times10^{-4}$ and $g^2/\lambda=2.0$, even though oscillons can be found for a wide range of these two parameters. The reason for this choice will become apparent in the next section. We note that with these values, the fundamental frequencies of oscillation near the minimum of the potential (at $\phi=0$ and $\psi=\pm 1$) are almost equal, with $\hat{m}_{\psi}=\sqrt{2}M$ and $\hat{m}_{\phi}\sim \sqrt{g^2/\lambda}M=\sqrt{2}M$.

\begin{figure}[htbp]
\includegraphics[width=0.49\linewidth]{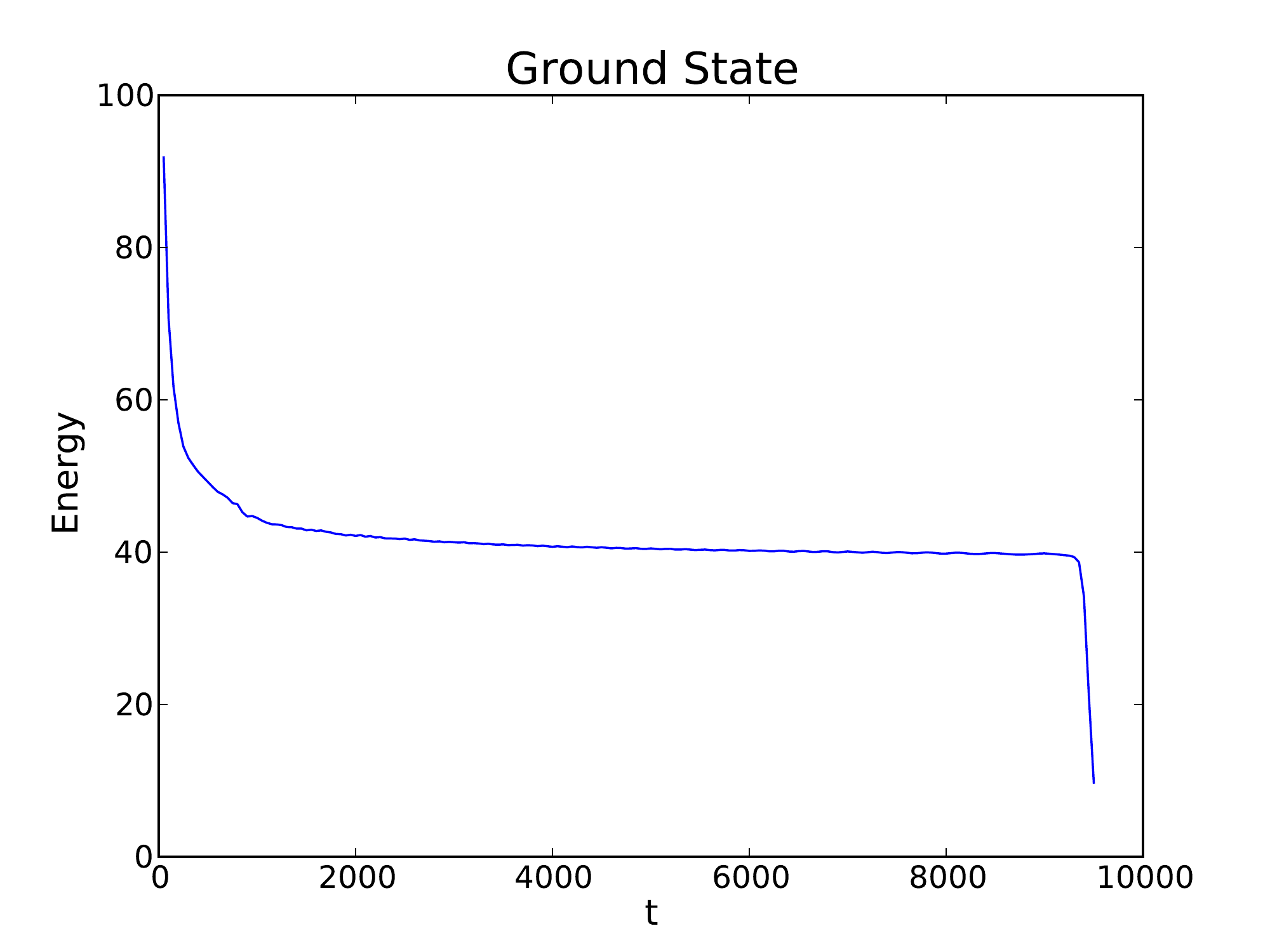}
\includegraphics[width=0.49\linewidth]{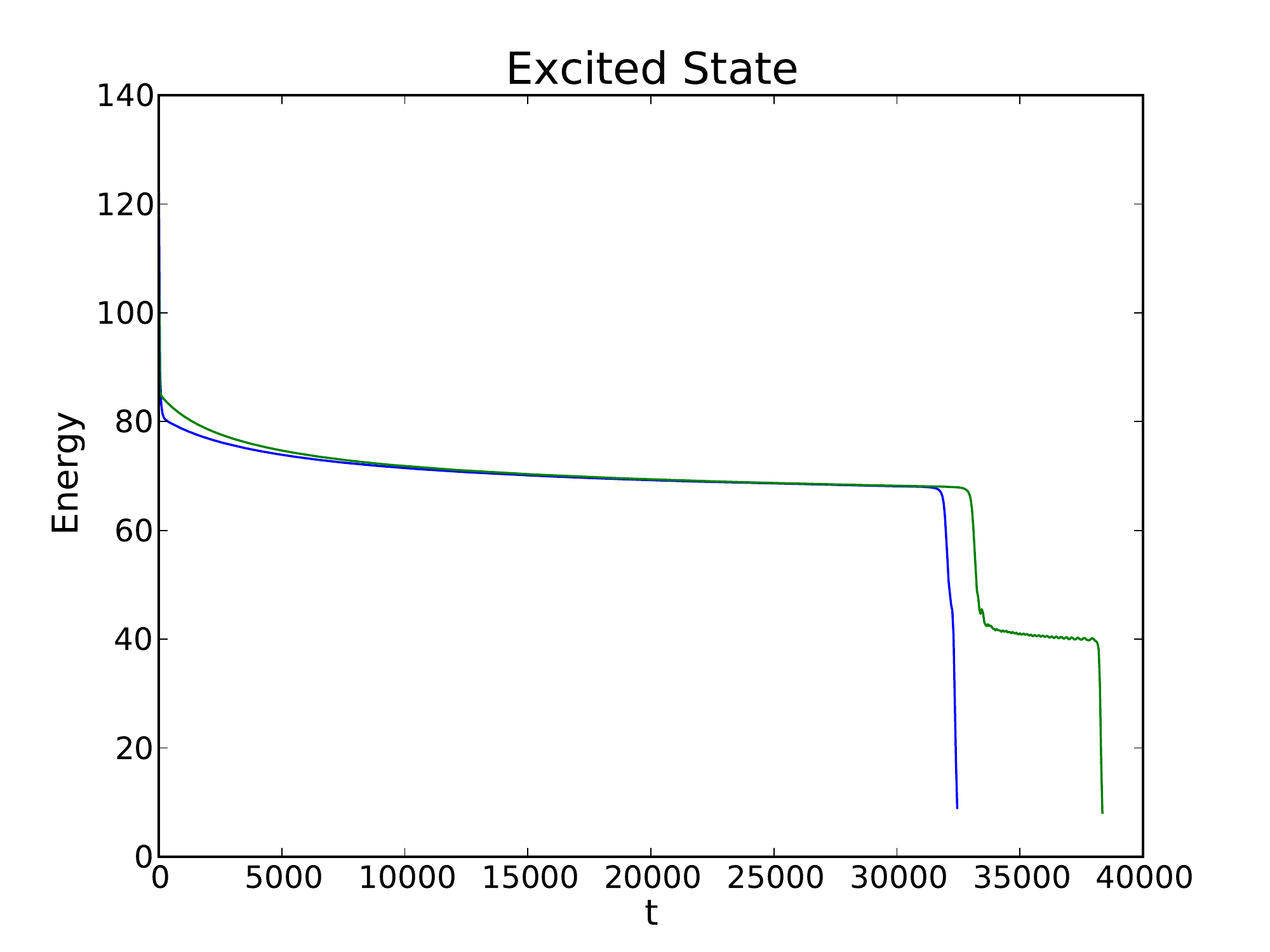} 
\caption{Oscillon energy versus time for three distinct classes of oscillons in our model. An oscillon in the ground state is plotted on the left. Two excited oscillon states are plotted on the right, one of which transitions to the ground state before decaying. Energy is given in units of $M/\lambda$ and time in units of $M^{-1}$. }
\label{osc_energy}
\end{figure}

We have solved the coupled nonlinear partial differential equations Eqs.~\ref{phi_eq_1d} and~\ref{psi_eq_1d} for a wide range of initial $R_{\psi}$ and $R_{\phi}$, the two free parameters in the initial Gaussian profiles of the fields, looking for combinations that lead to oscillon formation. Once formed, an oscillon's energy is operationally defined as the energy contained inside a sphere of radius of $R_{\rm shell}=10M^{-1}$ around the origin
\begin{equation}
 E_{\rm osc}=4\pi\int_0^{R_{\rm shell}}\left(\frac{1}{2}\left(\frac{\partial\phi}{\partial t}\right)^2+\frac{1}{2}\left(\frac{\partial\psi}{\partial t}\right)^2+\frac{1}{2}\left(\nabla\phi\right)^2+\frac{1}{2}\left(\nabla\psi\right)^2+V(\phi,\psi)\right)r^2dr.
\end{equation}
There are four possible outcomes: 1. The fields do not settle into an oscillon state and quickly radiate their initial energy to spatial infinity; 2. The fields settle into an oscillon state and reach an energy plateau at $E_{\rm osc}\sim43M/\lambda$, similarly to the single-field models studied in \cite{gleiser,copeland}. These oscillons survive for lifetimes ranging from $\sim3,000M^{-1}$ to $\sim 12,000M^{-1}$, depending on the initial Gaussian radii. We note that the longest-living single-field oscillon had a lifetime of $\sim 8,000M^{-1}$ \cite{copeland,fodor-honda}. This oscillon state will henceforth be called the \emph{ground state}; 3. The fields settle into an oscillon state with energy plateau $E_{\rm osc}\sim69M/\lambda$ and lifetimes up to $\sim33,000M^{-1}$. This configuration will be called the \emph{excited state}; 4. The fields settle into an excited oscillon state and then transition to the ground oscillon state where they remain for a while before decaying. In this case, we found that oscillons can remain localized for a total lifetime of $\sim40,000M^{-1}$, some five times longer than their cousins in single-field models. The energy evolution of the three classes of oscillons is plotted in Fig.\ref{osc_energy}. 

\begin{figure}[htbp]
\includegraphics[width=0.49\linewidth]{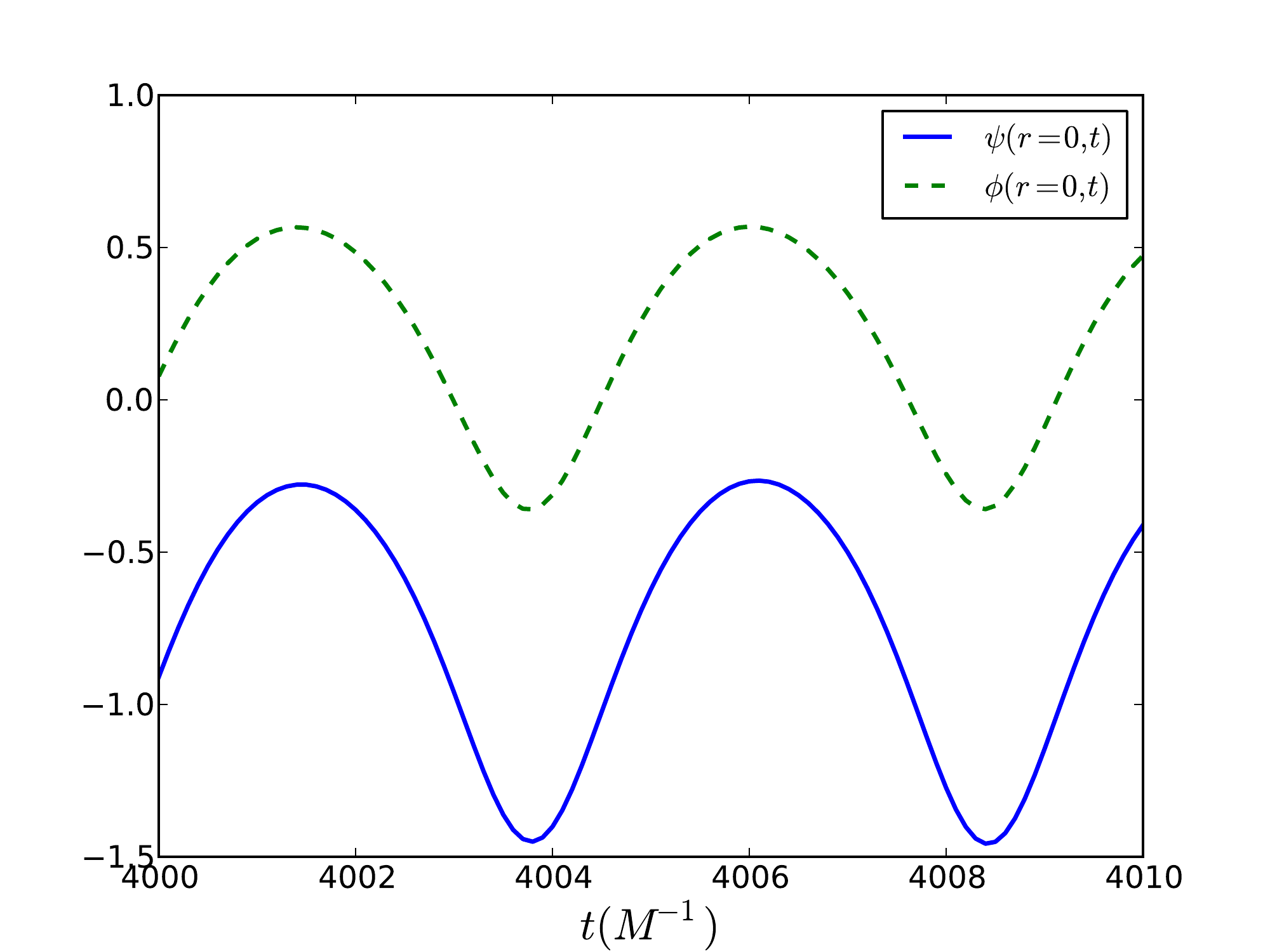}
\includegraphics[width=0.49\linewidth]{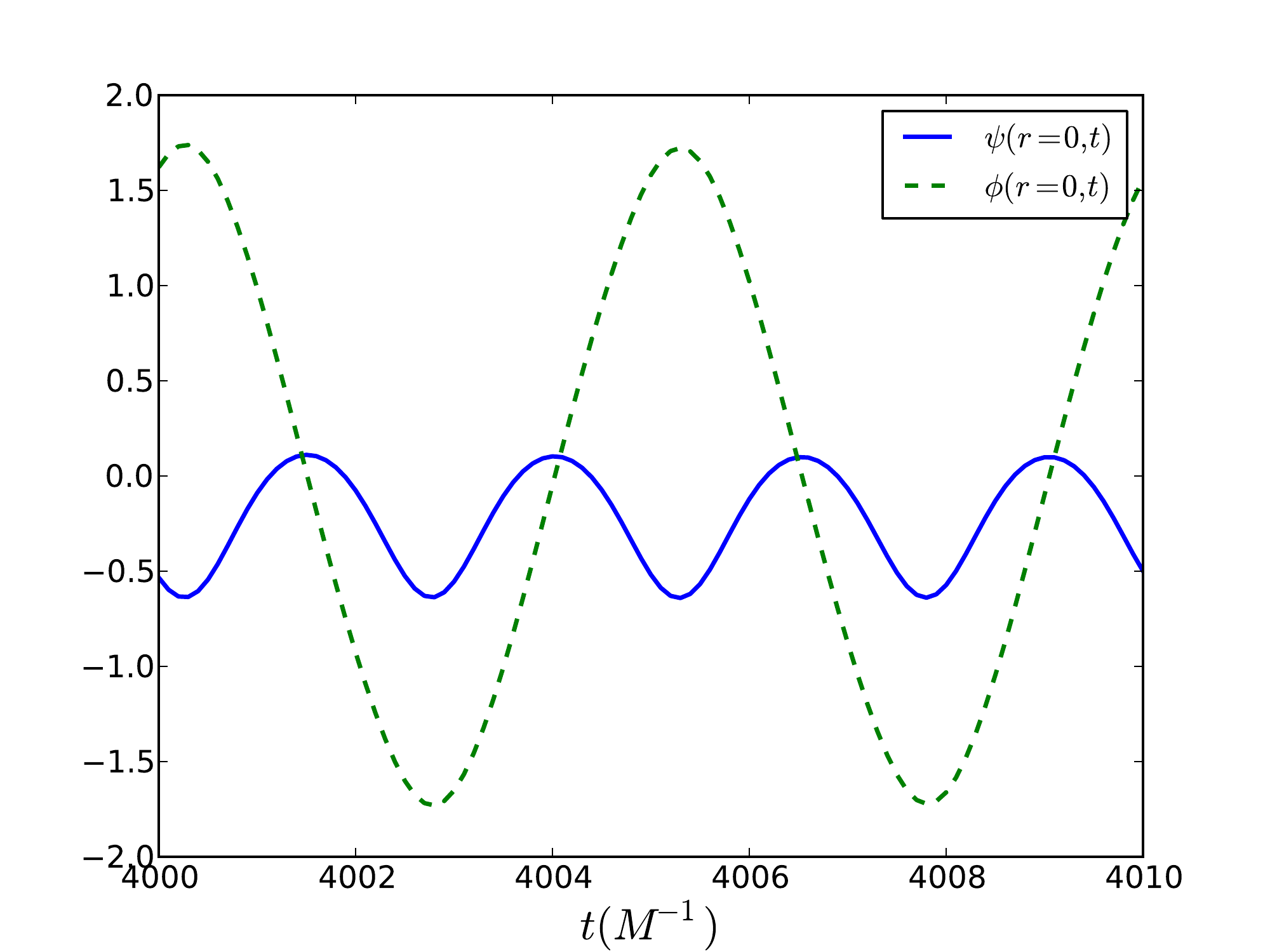}
\caption{Oscillation of the fields $\psi(r=0,t)$, $\phi(r=0,t)$ at the oscillon core, in the ground state (left) and in the excited state (right).}
\label{coreAmplitudes}
\end{figure}

The two oscillon states can be contrasted by looking at the behavior of the fields at the origin ($r=0$), as  shown in Fig.~\ref{coreAmplitudes}. In the ground state, the two fields coherently oscillate about their respective minima, maintaining an almost Gaussian profile with $R_{\psi}\sim3.25$ and $R_{\phi}\sim3.2$. The behavior of $\psi$ is similar to the one-field oscillon, leading to lifetimes enhanced by roughly $\sim 30\%$. (Since we have not performed an extensive analysis for different values of the two parameters $g^2/\lambda$ and $m^2/M^2$, these statements may be modified in the general case. It would be interesting to examine this question in more detail to see whether resonant behavior between the two fields may exist leading to much longer oscillon lifetimes.) 

\begin{figure}[htbp]
\includegraphics[width=0.49\linewidth]{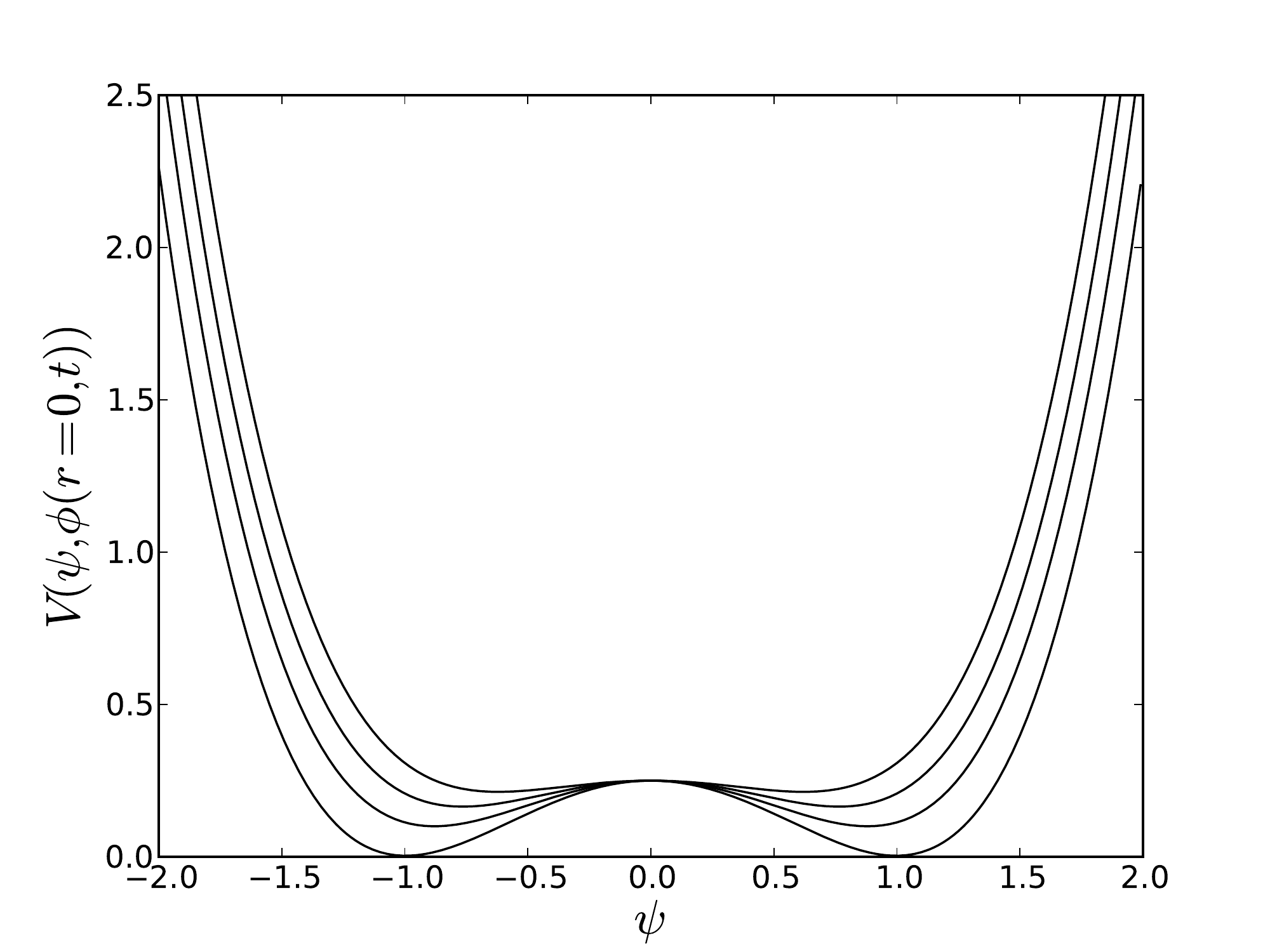}
\includegraphics[width=0.49\linewidth]{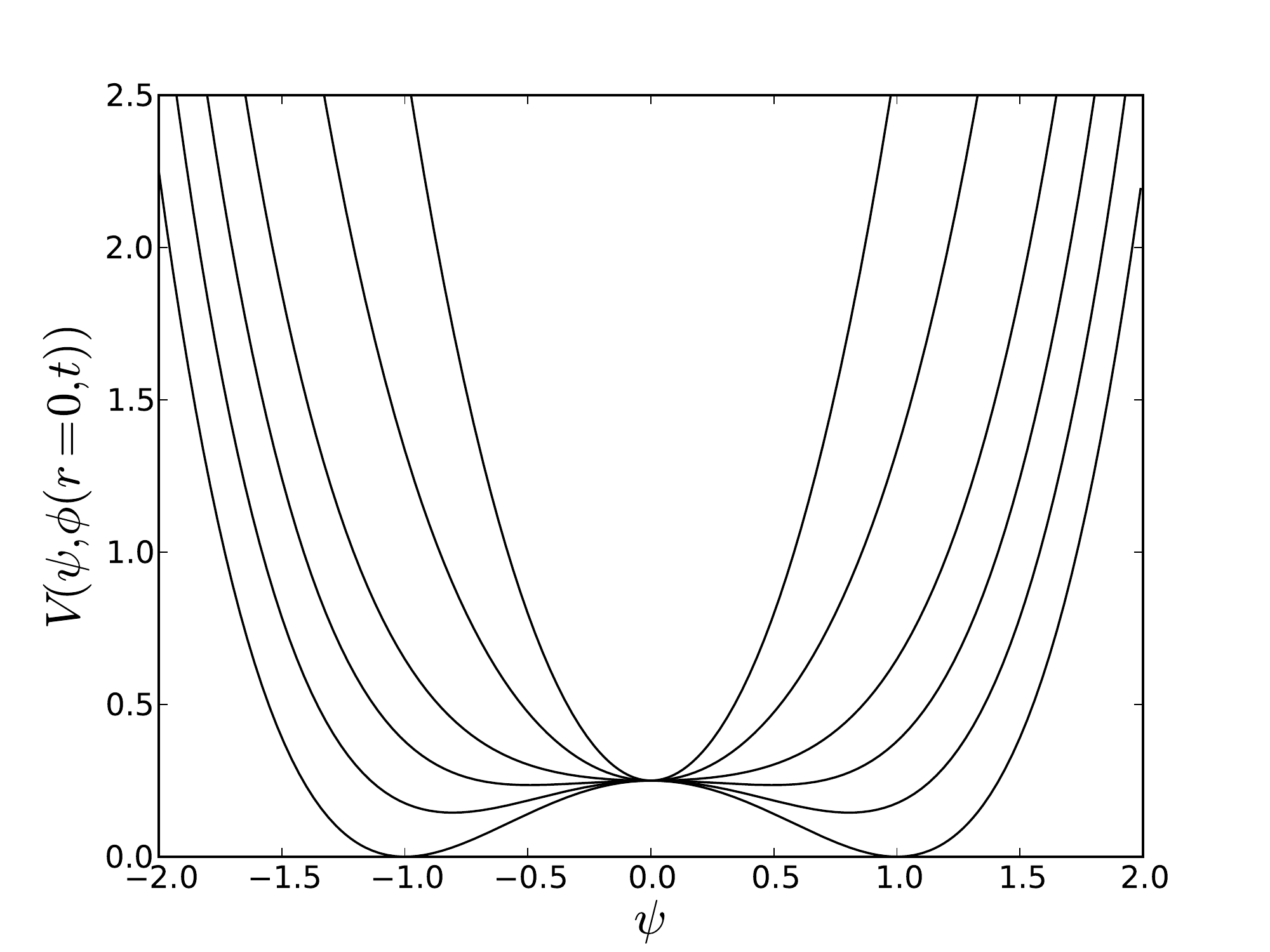}
\caption{Snapshots of the potential in the $\psi$ direction for the ground state (left) and for the excited state (right) at different times during a period in an oscillon's life. In the excited oscillon case, the amplitude of $\phi$ is large enough ($>\sqrt{2}/2$) to restore the symmetry and keep $\psi$ ``trapped'' without reaching its broken symmetric vacuum expectation value. In contrast, during the lifetime of the ground state oscillon the symmetry is never fully restored.}
\label{potential_range}
\end{figure}

In the excited oscillon state, the fields again reach a near-Gaussian profile with $R_{\psi}\sim2.0$ and $R_{\phi}\sim2.5$. However, $\phi$ oscillates about the minimum of its effective potential with large amplitude, and so the $g^2\phi^2\psi^2$ term in the potential of Eq.~\ref{hybrid_potential}, prevents $\psi$ from reaching its minimum: the field $\psi$ remains ``trapped'' and periodically returns to its symmetric state at $\psi=0$, which is why we identify this configuration the excited (or metastable) state denomination. This behavior leads to considerably longer oscillon lifetimes, suggesting that if large-amplitude oscillations of the field $\phi$ about its minimum could be maintained, oscillons may survive for arbitrarily long lifetimes. Figure~\ref{potential_range} illustrates how the large-amplitude oscillations of $\phi$ affect the potential for $\psi$. The amplitude of $\phi$ required to restore the symmetry has to satisfy
\begin{equation}
 \frac{\partial^2 V(\phi,\psi)}{\partial \psi^2}\bigg|_{\psi=0}>0,
\end{equation}
which gives $\phi>\sqrt{\lambda}/g= \sqrt{2}/2$ for our choice of parameters.

\begin{figure}[htbp]
\includegraphics[scale=0.5]{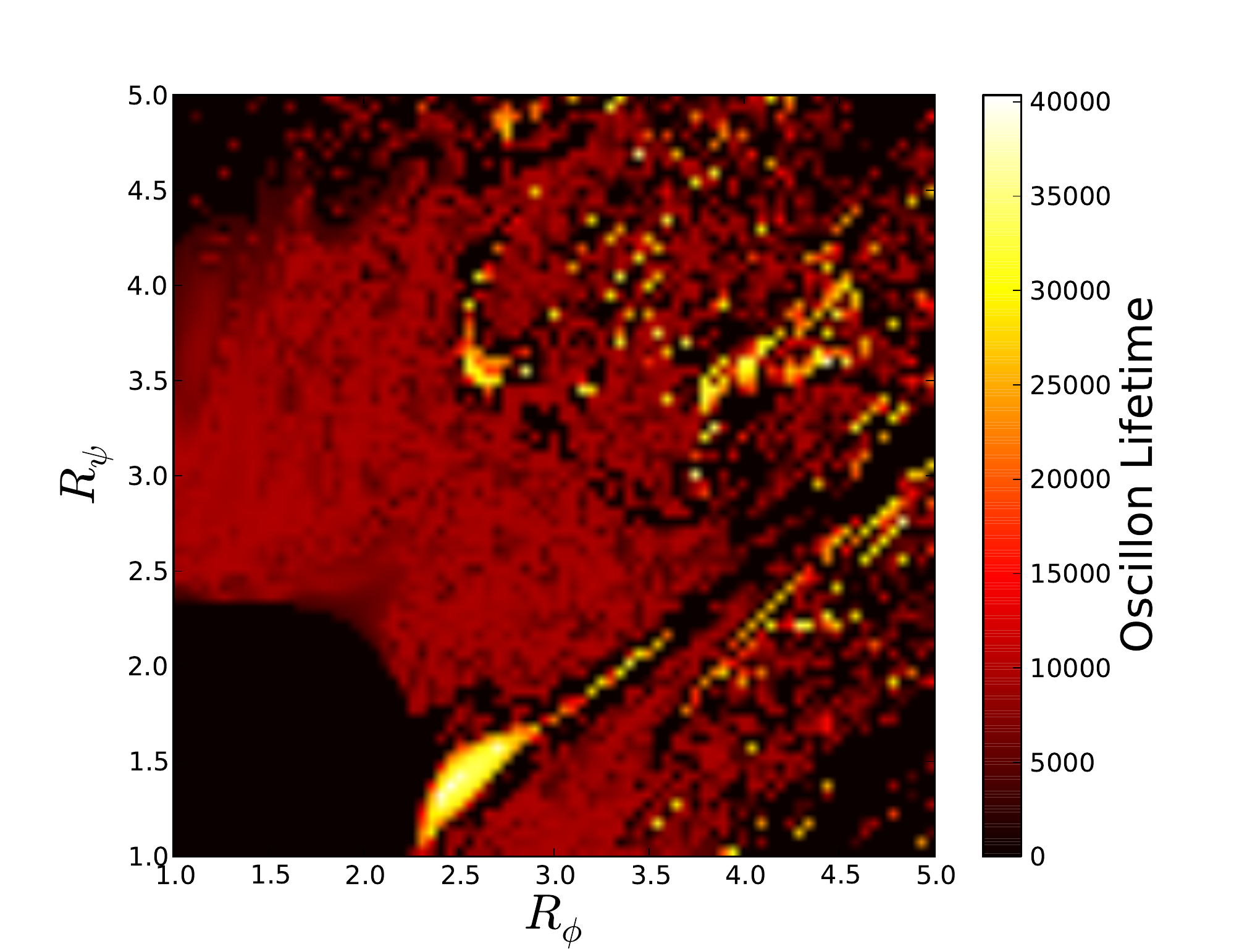}
\caption{Oscillon lifetime for different combinations of initial $R_{\psi}$ and $R_{\phi}$. Radii within the black regions form no oscillons. A color gradient towards white indicates the range of growing oscillon lifetimes.}
\label{lifetime_density_combined}
\end{figure}

\begin{figure}[htbp]
\includegraphics[width=0.49\linewidth]{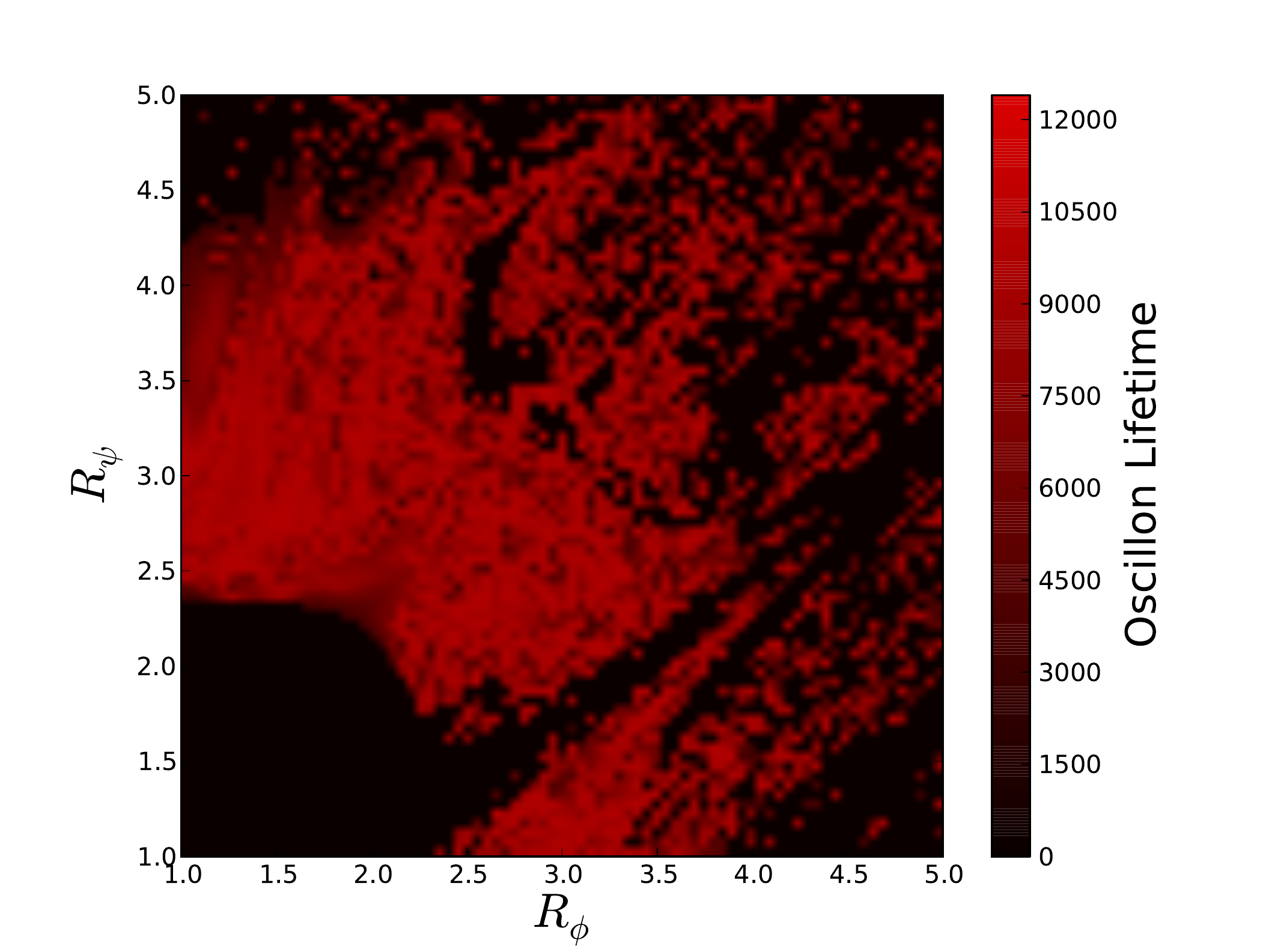}
\includegraphics[width=0.49\linewidth]{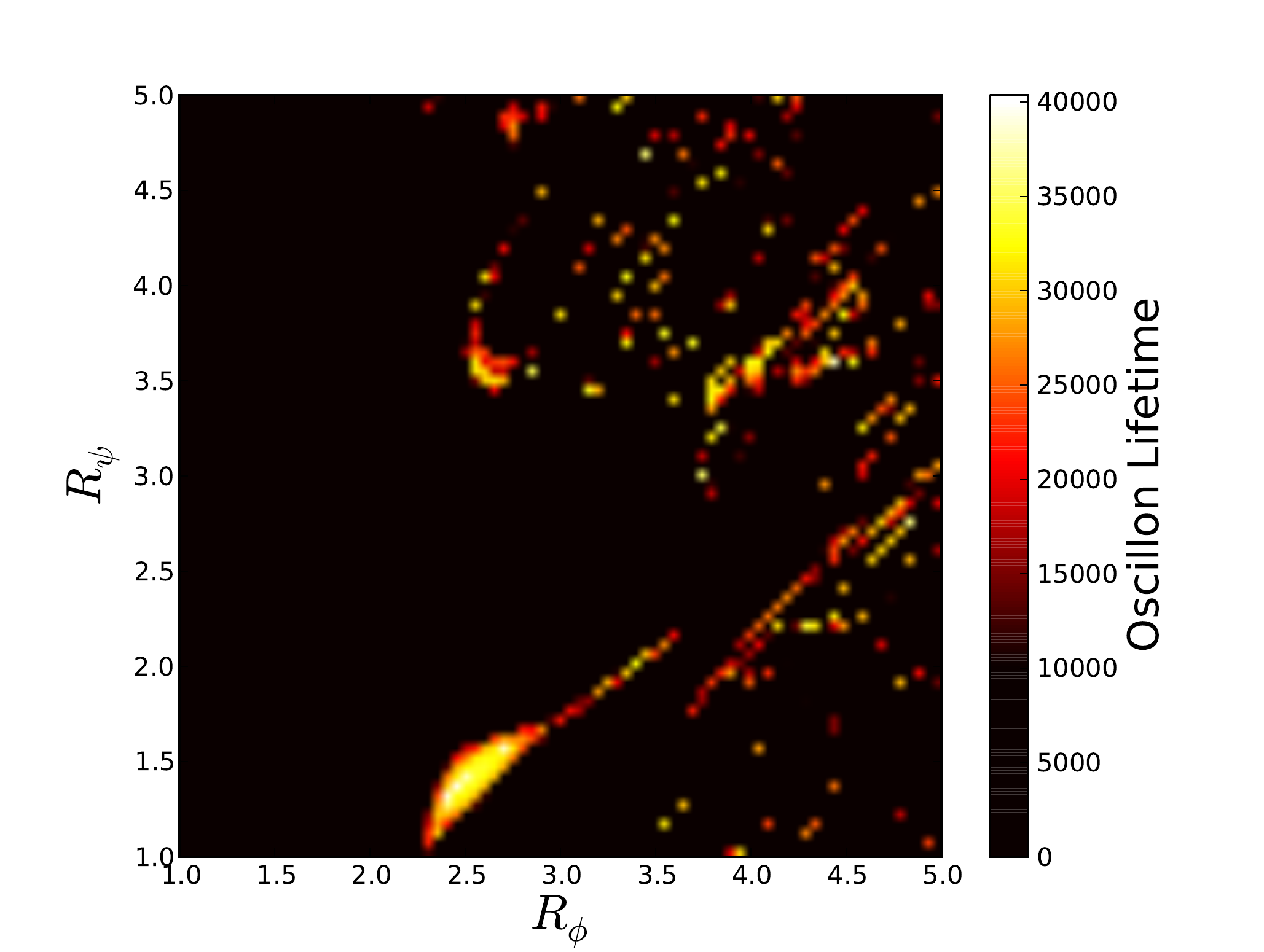}
\caption{Breakdown of Fig.~\ref{lifetime_density_combined} with ground state oscillons on the left and excited state oscillons on the right. Although excited state oscillons are rarer, they live considerably longer. There are also clear linear relationships between the radii, suggesting resonance patterns that remain to be explored.}
\label{lifetime_density_split}
\end{figure}

In Fig.~\ref{lifetime_density_combined}, we show how the existence and lifetime of oscillons depends on the choice of the initial Gaussian radii. We vary both radii from $1.0$ to $5.0$ in steps of $0.05$. The system portrays a subtly fine dependence on the initial conditions. Areas where the system does not settle into an oscillon are shown in black, whereas a nonblack color indicates oscillons with different lifetimes. In Fig.~\ref{lifetime_density_split} we separate the same data for ground state and excited state oscillons. 

For small radii, energy considerations alone clearly preclude oscillon formation. An obvious lower bound is $E(R_{\psi},R_{\phi})\gtrsim 43M/\lambda$, which lies in a band roughly given by $R_{\phi}^2 + R_{\psi}^2 \leq (1.5)^2$. However, we note that no oscillons form for a band roughly bounded by $R_{\phi}^2 + R_{\psi}^2 \leq (2.4)^2$. A similar lower bound on the initial radius that gives rise to oscillons was obtained for a single oscillon case \cite{copeland}. We can use the same approach to obtain an analogous result in the two-field model. First, write the fields as $\phi(r,t) = \phi_0(t)\exp[-r^2/R^2_\phi]$ and $\psi(r,t) = \psi_0(t)\exp[-r^2/R^2_\psi]-1$.
Integrating the Lagrangian density in the radial coordinate and varying the result with respect to the field amplitudes $\phi_0(t)$ and $\psi_0(t)$, we obtain two coupled ordinary nonlinear differential equations. Performing a small-amplitude expansion in both fields ($\phi=\phi_0 +\delta\phi$ and $\psi=\psi_0 +\delta\psi$) and keeping terms linear in the fluctuations, we can show, focusing on $\delta\psi$, that fluctuations will be unstable if 
\begin{equation}
\omega^2 \equiv \frac{3\sqrt{2}}{4}\psi_0^2 - \frac{4\sqrt{6}}{3}\psi_0+2+\frac{3}{R_{\psi}^2}+\frac{g^2/\lambda}{(1+R_{\psi}^2/R_{\phi}^2)^{3/2}}\phi_0^2 < 0,
\end{equation}
where $\phi_0$ is the amplitude of the $\phi$ field. These unstable fluctuations signal the existence of oscillons \cite{gleiser,copeland}. Given that $\omega^2$ is a parabola with minimum at $\psi_0^c = 8\sqrt{3}/9$, we find that for oscillons to exist, we must have
\begin{equation}
2 - \frac{16\sqrt{2}}{9} + \frac{3}{R_{\psi}^2} + \frac{g^2/\lambda}{(1+R_{\psi}^2/R_{\phi}^2)^{3/2}}\phi_0^2 < 0.
\end{equation}
The inequality is thus consistent with the black band of Fig. \ref{lifetime_density_combined}, since the contribution from the field $\phi$ never acts to decrease the minimum oscillon radius at $R_{\psi }= 2.42$. For $g^2=0$ we reproduce the result in Ref. \cite{copeland}, $R_{\psi }\geq 2.42$.

\section{Hybrid Inflation: Model And Constraints}
\label{sec:model}
We now investigate whether such two-field oscillons could play a role in cosmology. We turn our attention to the simplest realization of hybrid inflation \cite{linde_hybrid}, governed by the potential described in Eq.~\ref{hybrid_potential}. With a Friedmann-Robertson-Walker metric, the complete equations of motion for the two fields become

\begin{equation}
 \ddot{\phi}+3H\dot{\phi}-\frac{\nabla^2\phi}{a^2}=-(m^2+g^2\psi^2)\phi
\end{equation}
\begin{equation}
 \ddot{\psi}+3H\dot{\psi}-\frac{\nabla^2\psi}{a^2}=(M^2-g^2\phi^2-\lambda\psi^2)\psi,
\end{equation}
subject to the Friedmann equation,

\begin{equation}
 H^2=\frac{1}{3m_{\rm pl}^2}\left[\frac{1}{2}\dot{\phi}^2+\frac{1}{2}\dot{\psi}^2+\frac{1}{2}\frac{(\nabla\phi)^2}{a^2}+\frac{1}{2}\frac{(\nabla\psi)^2}{a^2}+V(\phi,\psi)\right],
\end{equation}
where $m_{\rm pl}$ is the reduced Planck mass.

For large values of $\phi$, the effective mass squared is positive for both $\phi$ and $\psi$ and the potential has the symmetry $\psi\leftrightarrow-\psi$ with a global minimum at $\phi=\psi=0$. For values of $\phi$ below the critical point $\phi_c=M/g$, the effective mass squared of $\psi$ becomes negative and symmetry breaking occurs, shifting the global minimum to $\psi_c=M/\sqrt{\lambda}$. Fast inflation is realized while $\phi$ slowly rolls down its potential with $\psi=0$, and lasts until $\phi$ drops below the critical point $\phi_c$, where symmetry breaking occurs and inflation ends. During inflation, the potential is $V(\phi,\psi=0)=M^4/4\lambda+m^2\phi^2/2$. Since $\phi$ is of the order $\phi_c=M/g$, for $m^2 \ll g^2M^2/\lambda$ the vacuum energy dominates the total energy density. The expansion rate during inflation is then well approximated by $H_{\rm inf}^2\sim M^4/(12\lambda m_{\rm pl}^2)$ \cite{bellido-linde}. In the limit $m\ll H_{\rm inf}$, the amplitude of the curvature perturbation spectrum can be calculated exactly in the case of hybrid inflation \cite{bellido-spectrum}

\begin{equation}
 {\cal P}_{\cal R}^{1/2}=\frac{1}{4\sqrt{48}\pi}\frac{g}{\lambda\sqrt{\lambda}}\frac{M^5}{m^2m_{pl}^3},
\end{equation}
and WMAP 7-year data constrain our parameters to satisfy

\begin{equation}
\frac{g}{\lambda\sqrt{\lambda}}\frac{M^5}{m^2m_{\rm pl}^3}\simeq4.29\times10^{-3},
\label{constraint}
\end{equation}
where we have used $\Delta_{\cal R}^2(k_0)=2.43\times 10^{-9}$ for $k_0=0.002\textrm{ Mpc}^{-1}$ \cite{WMAP}. In the following analysis we ensure the parameters satisfy this constraint.

\section{Hybrid Inflation: Simulations and Results}
\label{sec:results}
We now report on our numerical study of the dynamics of the two fields $\phi$ and $\psi$ after the end of inflation. We use the same rescalings as in Sec.~\ref{sec:1d}. With the dimensionless expansion rate $\tilde{H}=HM^{-1}$, the equations of motion become

\begin{equation}
 \ddot{\phi}+3H\dot{\phi}-\frac{\nabla^2\phi}{a^2} = -\left(\frac{m^2}{M^2}\phi+\frac{g^2}{\lambda}\psi^2\phi\right)
\label{phi_eq}
\end{equation}
\begin{equation}
 \ddot{\psi}+3H\dot{\psi}-\frac{\nabla^2\psi}{a^2} = \left(\psi-\frac{g^2}{\lambda}\phi^2\psi-\psi^3\right).
\label{psi_eq}
\end{equation}
The simulation space consists of a cube with comoving size $L$ and volume $V=L^3$ discretized on a regular lattice with spacing $\Delta x^i=\Delta r~(i=1,2,3)$. We simulate the initial conditions of the fields as quantum perturbations around a homogeneous value. We take the homogeneous value of $\phi$ to be $\phi_c=\sqrt{\lambda}/g$, the point that signals the end of inflation, while for $\psi$ we take $\psi=0$ just as its symmetry is about to be broken. We use periodic boundary conditions. To set up the initial conditions, we label both free fields' normal modes by $\mathbf{k}=(2\pi \mathbf{n}_i/L)$, where $\mathbf{n}=(n_x,n_y,n_z)$ and the $n_i$ are integers $n_i=-N/2+1\ldots N/2$.  Here, $N=L/\Delta r$ is the number of lattice points per side. Each free mode is described by a harmonic oscillator with frequency $\omega_k^2=(2\sin(k\Delta r/2)/\Delta r)^2+m_{\rm eff}^2$, where $k=|\mathbf{k}|$ and $m_{\rm eff}$ is the effective mass of each field given by Eq.~\ref{field_masses}. The initial conditions for the fields are then given by

\begin{eqnarray}
 \phi(\mathbf{r},t=0)&=&\phi_c+\frac{1}{\sqrt{V}}
\sum_\mathbf{k}\sqrt{\frac{1}{2\omega_k}}\left[\alpha_ke^{i
\mathbf{k}\cdot\mathbf{r}}+
\alpha_k^*e^{-i \mathbf{k}\cdot\mathbf{r}}\right],\nonumber\\
\dot{\phi}(\mathbf{r},t=0)&=&\frac{1}{\sqrt{V}}
\sum_\mathbf{k}\frac{1}{i}\sqrt{\frac{\omega_k}{2}}
\left[\alpha_ke^{i \mathbf{k}\cdot\mathbf{r}}-
\alpha_k^*e^{-i \mathbf{k}\cdot\mathbf{r}}\right],
\label{Eq:InitialCondPhi}
\end{eqnarray}

\begin{eqnarray}
 \psi(\mathbf{r},t=0)&=&\frac{1}{\sqrt{V}}
\sum_\mathbf{k}\sqrt{\frac{1}{2\omega_k}}\left[\alpha_ke^{i
\mathbf{k}\cdot\mathbf{r}}+
\alpha_k^*e^{-i \mathbf{k}\cdot\mathbf{r}}\right],\nonumber\\
\dot{\psi}(\mathbf{r},t=0)&=&\frac{1}{\sqrt{V}}
\sum_\mathbf{k}\frac{1}{i}\sqrt{\frac{\omega_k}{2}}
\left[\alpha_ke^{i \mathbf{k}\cdot\mathbf{r}}-
\alpha_k^*e^{-i \mathbf{k}\cdot\mathbf{r}}\right],
\label{Eq:InitialCondPsi}
\end{eqnarray}
where $\alpha_k$ is a random complex variable with phase distributed uniformly on $[0,2\pi)$ and magnitude drawn from a Gaussian distribution such that $\langle|\alpha_k|^2\rangle=\lambda/2$. Here we have expressed $k$ and $\omega_k$ in units of $M$ while $\Delta r$ has units of $M^{-1}$. The factor of $\lambda$ appears in our initial conditions because of our rescalings described above. The evolution of the expansion rate in dimensionless units is given by solving Eqs.~\ref{phi_eq} and \ref{psi_eq} together with Friedmann's equation

\begin{equation}
 H^2=\frac{1}{3}\frac{M^2}{m_{\rm pl}^2}\frac{\langle\rho\rangle}{\lambda},
\end{equation}
where $\langle\rho\rangle$ is the dimensionless spatially-averaged energy density. At each time step, we solve the three coupled equations for $\phi$, $\psi$, and for the scale factor $a(t)$, using $H=\dot{a}/a$ and $a(t=0)=1$.

\begin{figure}[htbp]
\includegraphics[scale=0.5]{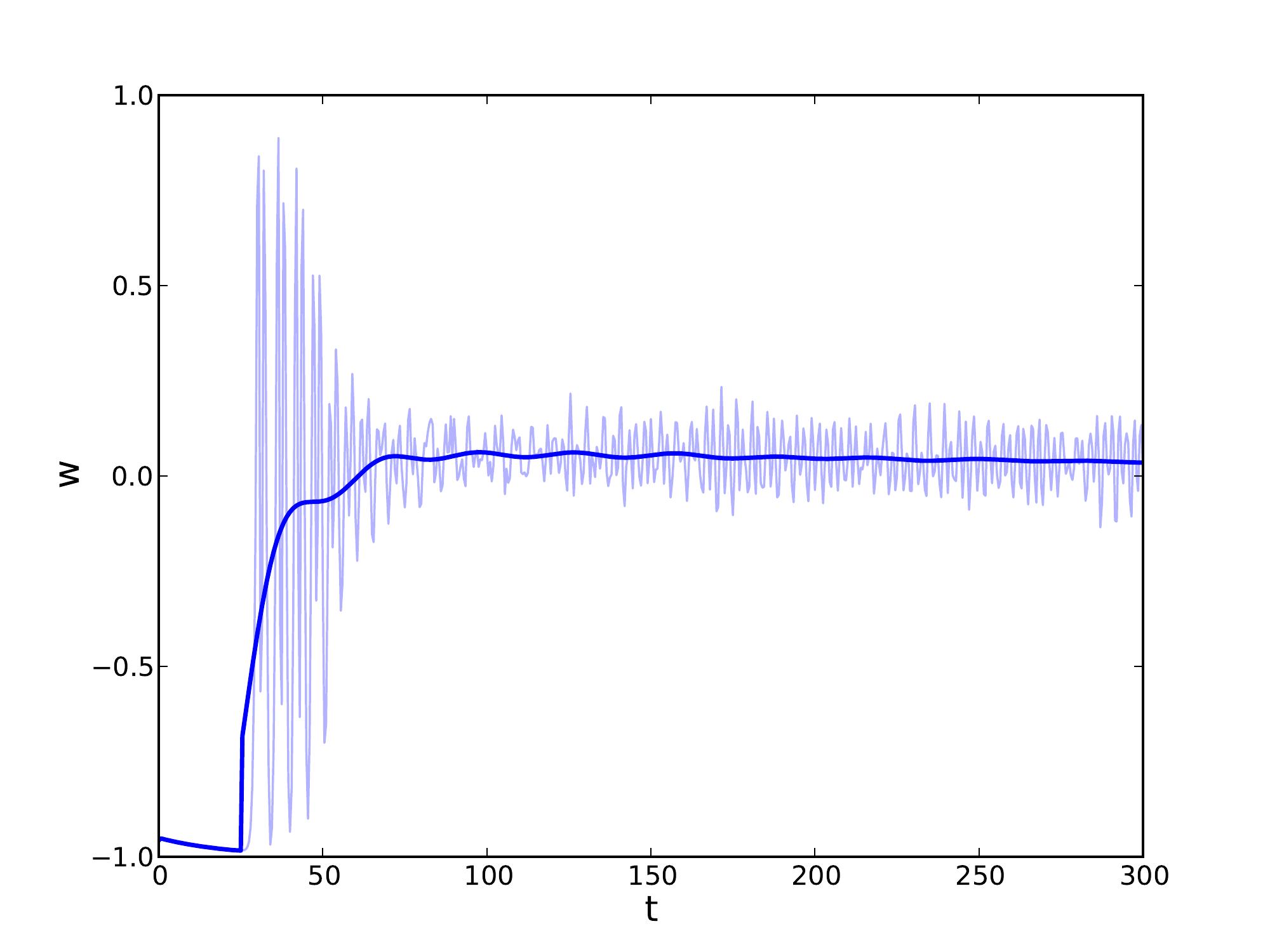}
\caption{The effective equation of state parameter $w\equiv\langle p\rangle/\langle\rho\rangle$ for a typical simulation, showing the transition from a near de Sitter expansion to a matter-dominated expansion.}
\label{eq_of_state}
\end{figure}

\begin{figure}[htbp]
\includegraphics[scale=0.5]{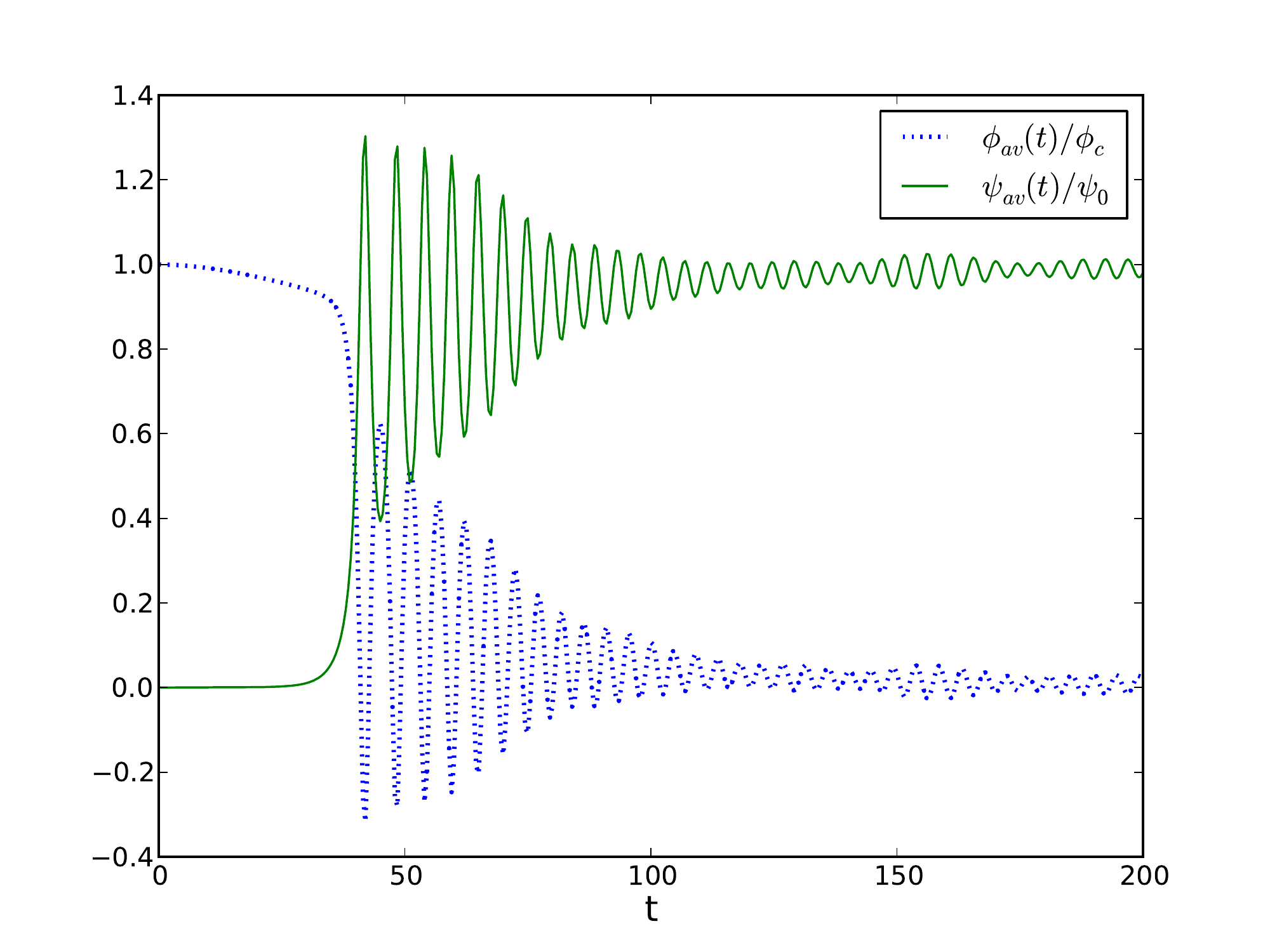}
\caption{Spatially-averaged evolution of both scalar fields at the end of hybrid inflation with $\phi_c=M/g$ and $\psi_0=M/\sqrt{\lambda}$.}
\label{average_fields}
\end{figure}

We discretize the equations of motion using second-order space derivatives with lattice spacing $\Delta r_0=0.05M^{-1}$, second-order time derivatives with $\Delta t=0.01M^{-1}$, in a box with $256^3$ lattice points. The effective equation of state parameter $w\equiv\langle p\rangle/\langle\rho\rangle$ is plotted in Fig.~\ref{eq_of_state}. Since $w$ naturally undergoes large fluctuations, we use a window function to uncover its underlying behavior. As can be seen in the figure, initially the Universe undergoes a near de Sitter expansion with constant expansion rate $H^2\sim H_{\rm inf}^2= M^4/(12\lambda m_{\rm pl}^2)$ and $w\lesssim -1$. It then quickly transitions into a matter-dominated Universe which, as expected from the coherent oscillations of the inflaton about the origin, behaves like pressureless dust, with $w=0$ and $H(t)\propto 2/(3t)$. In this phase, the two fields oscillate nonlinearly about their respective minima with amplitudes that depend on their relative effective masses. (See, for example, Fig.~\ref{coreAmplitudes}, in particular the left plot). Although we have not searched extensively, we were thus far unable to find excited state oscillons in a cosmological setting. The difficulty is mostly practical, since excited states require large initial oscillations in the $\phi$ field and this makes simulations harder to carry out. The spatially-averaged behavior of the two fields in a typical simulation can be seen in Fig.~\ref{average_fields}. Throughout our simulations, we set $M=7\times10^{-5} m_{\rm pl}$ and $\lambda = 3\times10^{-6}$, and we investigate the evolution of the system for different values of the ratio $g^2/\lambda$ by changing $g$. At the same time, we adjust the value of $m$ to be consistent with the WMAP constraint of Eq.~\ref{constraint}. 

We identify three regimes in $g^2/\lambda$. For the weakly-coupled system with $g^2/\lambda \ll 1$, $m_{\phi}\ll m_{\psi}$ and most of the energy of the system is transferred to $\phi$. As a consequence, after the two fields reach the minimum of the potential, they start oscillating with very different amplitudes. While $\phi$ oscillates with large amplitude about $\phi=0$, the oscillations of $\psi$ about its minimum at $\psi_0=M/\sqrt{\lambda}$ have very small amplitude. As the system evolves, $\phi$ loses energy to the expansion and is left with uniform fluctuations throughout the simulation volume. In the meantime, $\psi$ uniformly settles at its minimum and no oscillons are formed. This is the regime usually studied during tachyonic preheating \cite{felder-preheating}.

For $g^2/\lambda \gtrsim 1$ however, $m_{\phi}\gtrsim m_{\psi}$ and the amplitude of oscillations in $\psi$ become comparable to and even larger than those in $\phi$. These nonlinear, large-amplitude fluctuations lead to the formation of oscillons with an approximate radius of  $R_{\rm osc}\lesssim 5M^{-1}$, a near-Gaussian profile (just as the ground state oscillons of Sec.~\ref{sec:1d}) and energy $E_{\rm osc}\sim43M/\lambda$. In spite of the red-shifting of field modes due to the expansion, these objects remain coherent throughout our simulation which goes to times longer than $800M^{-1}$. This time scale should be compared with the Hubble time during their formation, of order $H^{-1}\sim200M^{-1}$. Even though tracking the oscillons for much longer times in an expanding Universe remains challenging numerically, our results indicate that oscillons act cosmologically as stable coherent objects, as has been conjectured in Ref.~\cite{copeland}.

\begin{figure}[htbp]
\includegraphics[scale=0.5]{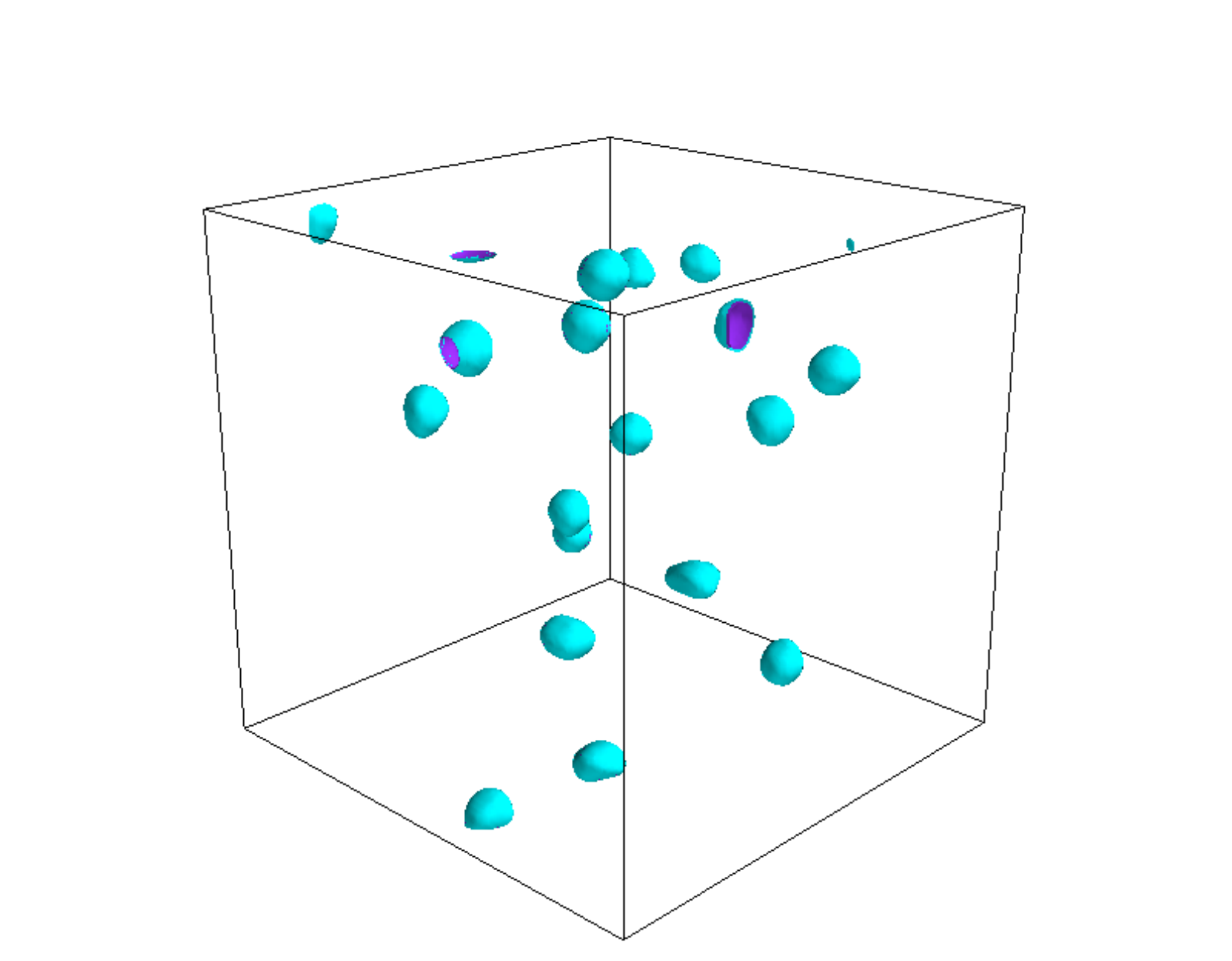}
\caption{Energy density of the system after oscillons have formed for $g^2/\lambda=2$. Two isosurfaces are shown, one at $10$ times the average energy of the box (light blue) and one at $12$ times the average energy (purple). The system has been evolved until the scale factor is $a(t_f)=7$. The oscillons come to rest in the comoving frame and stay localized for the duration of the simulation. We have used $M=1.7\times10^{14}\textrm{ GeV}$, $m=1.0\times10^{12} \textrm{ GeV}$ and $H=2\times10^{12} \textrm{ GeV}$. The oscillons have approximate energy $E_{\rm osc}\simeq 25\times10^{20}\textrm{ GeV}$, confined to a radius of approximately $6\times 10^{-26}$ cm. A visualization of a typical run can be seen at \cite{simulation}.} 
\label{oscillon_snapshot}
\end{figure}

Finally, for $g^2\gg \lambda$, $\phi$ can be neglected in the equation of motion for $\psi$ and our system is equivalent to the one studied in Ref.~\cite{3d_osc_temperature}, which demonstrated the existence of cosmological oscillons in a 3d, single-field model with a symmetric double-well potential. (Note, however, that in that previous investigation the expansion rate was kept constant for the duration of the simulation, while here we solve the coupled Friedmann-Klein-Gordon system self-consistently.) A snapshot of the energy density after oscillons have formed is shown in Fig.~\ref{oscillon_snapshot}.  

\begin{figure}[htbp]
\includegraphics[scale=0.5]{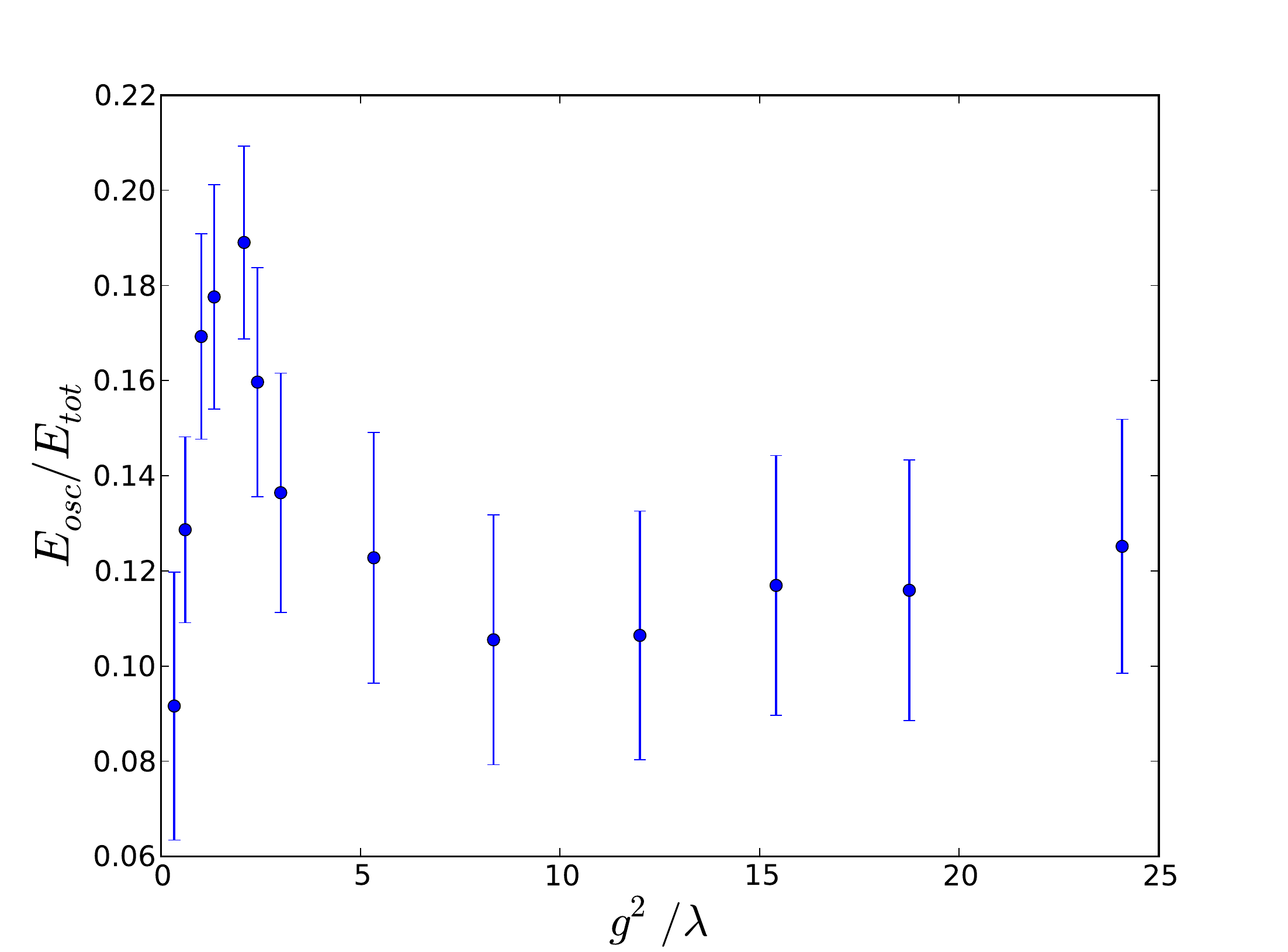}
\caption{Fraction of the energy in oscillons as a function of the coupling ratio $g^2/\lambda$. Each point represents an average over 50 runs with error bars giving one standard deviation.} 
\label{rho_osc}
\end{figure}

Given that oscillons emerge at the end of hybrid inflation, it is important to compute their contribution to the total energy of the Universe. We have performed simulations until $a(t)=7$ for several values of $g^2/\lambda$, measuring the energy of an oscillon by locating its core and integrating the energy around it within a sphere of radius $R_{\rm shell}=5M^{-1}$. We have verified that our results remain invariant using slightly higher or lower values of $a(t_f)$ and $R_{\rm shell}$. Since our initial conditions depend on random amplitudes for the fields, we average over multiple runs for each value of $g^2/\lambda$. Our results are shown in Fig.~\ref{rho_osc}. These simulations have been performed using a parallel multicore system and a new GPU-enhanced code, which will be described in a forthcoming publication \cite{gpu}. Several features are immediately apparent in Fig.~\ref{rho_osc}. As we remarked before, for $g^2/\lambda<1$, $\psi$ oscillates with small amplitude about its minimum, and thus cannot probe the nonlinear part of the potential, which precludes the formation of oscillons in these models \cite{gleiser,copeland}. Indeed, for $g^2/\lambda\lesssim0.1$ we do not see oscillons emerge and find that their contribution to the total energy is negligible. On the other hand, for $g^2/\lambda\gtrsim3$, the oscillations in $\phi$ are too insignificant to affect the evolution of $\psi$ and we find a contribution for oscillons to the total energy of about $10\%$, roughly independently of $g^2/\lambda$. For $g^2/\lambda\sim2$, we find a significant peak in the fraction of the total energy contributed in oscillons, with a maximum at $\sim20\%$. As we explain next, this peak is the result of parametric resonance induced by the coupling between the two fields.

We decompose $\psi$ as $\psi(\textbf{x},t)=\psi_{av}(t)+\delta\psi(\textbf{x},t)$, where $\psi_{av}$ is the volume averaged field. Linearizing Eq.~\ref{psi_eq} with respect to $\delta\psi(\textbf{x},t)$ and taking the Fourier transform while treating $\phi$ as a function of time $\phi=\phi(t)$, we obtain (for $k>0$)
\begin{equation}
\delta\ddot{\psi}(k,t) + 3H\delta\dot{\psi}(k,t)+
\left(\frac{k^2}{a(t)^2} + 3\psi_{av}^2+\frac{g^2}{\lambda}\phi^2(t)-1\right)\delta\psi(k,t)=0.
\label{Eq:LinearEq}
\end{equation}
The cosmic expansion is important only in the first few oscillations and quickly becomes insignificant during oscillon production. It does, however, affect the physical modes that enter Eq.~\ref{Eq:LinearEq}, since they get red-shifted as $k_{\rm phys}=k/a(t)$. We could have taken the expansion into account by redefining $\chi=a^{3/2}\psi$, but since in the post-inflationary regime the scale factor goes as $a(t)\propto t^{2/3}$,   the friction term due to the cosmic expansion will become negligible. As we will see, this choice will simplify our analysis while still incorporating the dominant effects from the expansion. 

After the fields have reached their minima, their near-sinusoidal oscillations can be approximated as $\psi_{av}=\bar{\psi}+\Psi\cos{\omega_{\psi}t}$ and $\phi(t)=\Phi\cos{\omega_{\phi}t}$, where $\bar{\psi} \sim \psi_0=M/\sqrt{\lambda}$ and $\omega_{\psi}=\sqrt{2}M$ and $\omega_{\phi}=g/\sqrt{\lambda} M$ will be the fundamental frequencies of oscillation. Here, $\Phi$ and $\Psi$ denote the amplitude of oscillation about the minimum for each field, which depend on the value of $g^2/\lambda$. Inserting these into Eq.~\ref{Eq:LinearEq} we get
\begin{equation}
 \delta\ddot{\psi}(k,t)+
\left(\frac{k^2}{a(t)^2} + 3\bar{\psi}^2-1+6\bar{\psi}\Psi\cos{\omega_{\psi}t}+3\Psi^2\cos^2{\omega_{\psi}t}+\Phi^2\cos^2{\omega_{\phi}t}\right)\delta\psi(k,t)=0,
\label{Eq:LinearEq2}
\end{equation}
which we rewrite as
\begin{equation}
 \delta\ddot{\psi}(k,t)+\omega_0^2
\left(1+h\cos{\omega_{\psi}t}+\epsilon\cos^2{\omega_{\psi}t}+\delta\cos^2{\omega_{\phi}t}\right)\delta\psi(k,t)=0,
\label{Eq:LinearEq3}
\end{equation}
with $\omega_0^2=\frac{k^2}{a(t)^2} + 3\bar{\psi}^2-1$, $h=6\bar{\psi}\Psi/\omega_0^2$, $\epsilon=3\Psi^2/\omega_0^2$  and 
$\delta=\Phi^2/\omega_0^2$. For each wave number $k$ Eq.~\ref{Eq:LinearEq3} has effectively only one free parameter because all quantities that appear are either constant or depend on the value of $g^2/\lambda$.

\begin{figure}[htbp]
\includegraphics[scale=1.0]{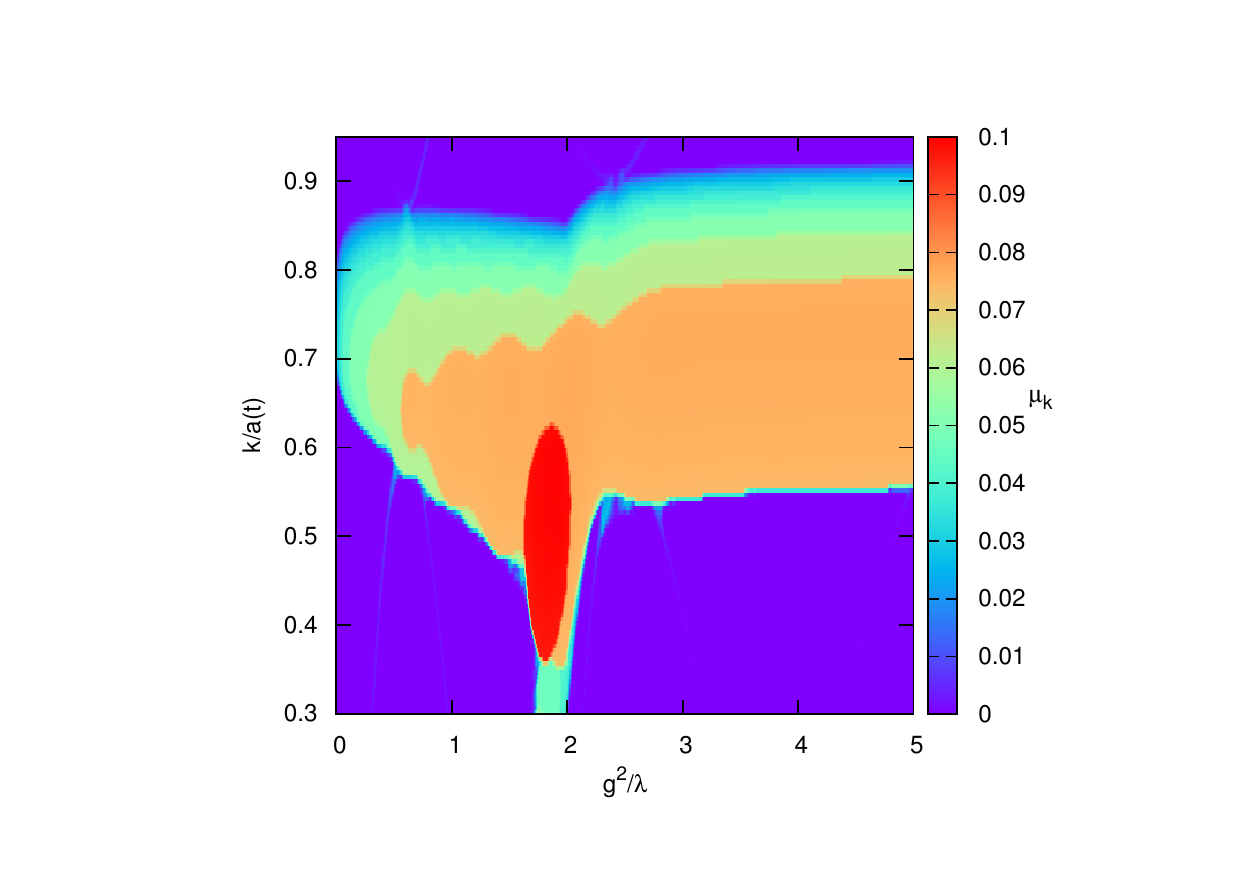}
\caption{Floquet exponent $\mu_k$ for solutions of Eq.~\ref{Eq:LinearEq3} with different values of $g^2/\lambda$ and $k/a(t)$. } 
\label{floquet}
\end{figure}

We would like to solve the above equation analytically and look for solutions of the form $\delta\psi(k,t)\propto e^{\mu_kt}P(t)$, where $\mu_k$ is real and positive and $P(t)$ is a periodic function, to see the effects of parametric resonance on modes $k/a(t)$. However, standard Floquet analysis \cite{landau-lifshitz} requires the values of $\delta$, $\epsilon$ and $h$ to be much smaller than unity, which is not the case in this model: $h$, in particular, can be larger than unity. We hence solve Eq.~\ref{Eq:LinearEq3} numerically, looking for solutions of the form $\delta\psi(k,t)\propto e^{\mu_kt}P(t)$ and identifying the Floquet exponent $\mu_k$. We investigate values of $0\leq g^2/\lambda \leq5$ and $0.3M\leq k/a(t)\leq1.0M$. For each value of $g^2/\lambda$ we obtain ensemble averages for the values of $\Phi$ and $\Psi$ from the runs of Fig.~\ref{rho_osc}, which are then used to solve Eq.~\ref{Eq:LinearEq3}. Our results are shown in Fig.~\ref{floquet}. Oscillons of radius $R_0$ are characterized by oscillations with wave number $2/R_0$, where oscillon radii range from $2.5M^{-1}$ to $3.5M^{-1}$ \cite{gleiser,copeland}. And so, as previously also shown in Ref. \cite{gleiser-howell}, their formation requires the excitation of modes with $0.57M<k/a(t)<0.8M$. In cosmological simulations, oscillons form most efficiently from initial overdensities of radii $4-6M^{-1}$, which correspond to a range of wave numbers $0.33M<k/a(t)<0.5M$. From Fig.~\ref{floquet}, we see that this is precisely within the range of wave numbers that undergoes largest exponential growth, corresponding to the oval shaped region around $g^2/\lambda \sim 2$ (colored in red). These overdensities subsequently radiate their excess energy, shrinking slightly to condense into oscillon configurations of smaller radii, within the known range of $2.5M^{-1}$ to $3.5M^{-1}$ noted above. For small values of $g^2/\lambda$, few modes are amplified and only weakly, explaining the left part of Fig.~\ref{rho_osc}. On the other hand, for $g^2/\lambda\geq3.0$, modes with $0.55M<k/a(t)<0.8M$ are significantly amplified, giving rise to a considerable number of oscillons. Note also the near-constant bands of the amplified modes for these values of $g^2/\lambda$, which clearly correlates with the flat plateau in energy in Fig.~\ref{rho_osc}. (As we remarked above, for values of $g^2/\lambda>3.0$, $\phi$ oscillates with small and eventually negligible amplitude, while $\psi$ reaches a maximum amplitude of oscillation.) Most importantly, near $g^2/\lambda\lesssim 2$, modes with $k/a(t)\sim0.4M$ undergo large amplification leading to copious production of oscillons. This explains the resonant peak in Fig.~\ref{rho_osc} for the contribution of oscillons to the total energy, $E_{\rm osc}/E_{\rm tot}$.

\section{Conclusions and Outlook}

We have investigated the nonlinear dynamics of hybrid inflation cosmological models featuring two coupled scalar fields: the inflaton $\phi$, with a quadratic potential, and a waterfall field $\psi$, with a double-well potential. The fields couple quadratically with strength $g^2$ \cite{linde_hybrid, bellido-linde, basset}. Exploring the model first in static Minkowski space, we have found that even without cosmic expansion, the model has very rich dynamical features. We have shown that it supports the existence of long-lived, spherically-symmetric, coherent time-dependent structures, which we described as a new class of oscillons \cite{gleiser, copeland}. These objects appear in two possible states, which we have called ground and excited. In some cases, the excited state was seen to relax into the ground state, where it persisted before finally decaying. The lifetimes of this new class of two-field oscillons were measured to be five times longer than their single-field counterparts. Given that our main focus in this work is the cosmological relevance of oscillonlike structures, we have not provided a detailed analysis of their properties, although their enhanced lifetime is likely a consequence of resonant behavior. We believe they are an excellent laboratory in which to investigate the effects of resonances in coupled-field models able to support localized, coherent (solitonlike) time-dependent field configurations, a topic worthy of further study in its own right.

Coupling the two-field model to the cosmological expansion, we have solved the Friedmann-Klein-Gordon system of three coupled equations in three spatial dimensions. We started the simulation near the end of inflation, as the inflaton field $\phi$ breaks the symmetry of the $\psi$ field, which relaxes to its global minimum. We note that the fields were initialized as Gaussian random variables, but remarkably self-organized to generate a network of long-lived oscillons as the fields oscillate about their minima. For certain choices of the couplings, we have shown that the long-lived structures persist for at least four cosmological expansion times. There were three regimes, determined by the value of the ratio of couplings, $g^2/\lambda$. The most interesting results were found for $g^2/\lambda \geq 1$, when oscillons were formed. Their contribution to the energy density of the Universe can be as much as $20\%$, in a renormalizable model with realistic couplings that is consistent with WMAP 7-year data. Finally, we explained the formation of oscillons and their substantial contribution to the total energy of the Universe as a consequence of parametric resonance.

There are many interesting open questions to explore. Apart from studying the dynamics of the two-field model in static space in more detail than in Sec. \ref{sec:1d}, it would be interesting to examine how our results would vary if we included couplings to other fields in both Minkowski and expanding spacetimes. In particular, even though the oscillations of $\phi$ are supposed to generate the reheating of the Universe and its transition to a radiation-dominated expansion, we have not incorporated the couplings to other fields that would initiate a radiation-dominated era. Hence, we found a transition to a matter-dominated cosmos instead. What will be the role of oscillons during reheating? If previous results are an indication, oscillons will act as bottlenecks for energy equipartition, delaying the approach to equilibrium \cite{gleiser-howell}. In this case, they or other possible time-dependent coherent structures will strongly influence the value of the reheating temperature, providing a key constraint to our understanding of the physics of the early Universe. The fact that oscillons may contribute a sizable fraction of the total energy in the Universe suggests that they could have played a role in baryogenesis. While baryons produced before inflation are inflated away or, if produced after, are wiped out as the Universe equilibrates, oscillons remain out of equilibrium.  It would interesting to examine if the time-scale associated with the oscillations of the scalar fields in oscillons, or even their lifetime itself, may be used as a mechanism to generate a baryon asymmetry. Work along these lines is currently in progress.

\section{Acknowledgements}
We thank Research Computing at Dartmouth College for access to their facilities and Susan Schwarz for technical support. MG was supported in part by National Science Foundation Grant No. PHY-0653341. NG was supported in part by National Science Foundation Grant No. PHY-0855426. NS was supported by the William H. Neukom 1964 Institute for Computational Science at Dartmouth College.


\begin{thebibliography}{99}

\bibitem{guth} A. H. Guth, Phys. Rev. D \textbf{23}, 347 (1981).

\bibitem{basset} B. A. Bassett, S. Tsujikawa and D. Wands, Rev. Mod. Phys. \textbf{78}, 537 (2006).

\bibitem{bbn} R. A. Alpher, H. Bethe and G. Gamow, Phys. Rev. \textbf{73}, 803 (1948).

\bibitem{kolb} E. Copeland, E. W. Kolb, A. R. Liddle, and J. E. Lidsey, Phys. Rev. D{\bf 48}, 2529 (1993); A. R. Liddle and M. S. Turner, Phys. Rev. D{\bf 50}, 758 (1994); Erratum-ibid. D{\bf 54}, 2980 (1996); B. A. Powell and W. H. Kinney, JCAP 0708, 006 (2007).

\bibitem{adshead} P. Ashead and R. Easther, JCAP 0810, 047 (2008).

\bibitem{linde_hybrid} A.D. Linde, Phys. Lett. \textbf{B259}, 38, (1991), Phys. Rev. D \textbf{49}, 748 (1994).

\bibitem{bellido-linde}J. Garcia-Bellido, A. Linde, Phys. Rev. D \textbf{57}, 6075-6088, (1998).

\bibitem{felder-preheating} G. Felder, J. Garcia-Bellido, P. B. Greene, L. Kofman, A. D. Linde and I. Tkachev, Phys. Rev. Lett \textbf{87}, 011601 (2001). 

\bibitem{copeland-preheating} E. J. Copeland, S. Pasoli and A. Rajantie, Phys. Rev. D \textbf{65}, 103517 (2002). 

\bibitem{mcdonald} J. McDonald, Phys. Rev. D \textbf{66}, 043525 (2002), M. Broadhead and J. McDonald, Phys. Rev. D \textbf{72}, 043519 (2005).

\bibitem{bogol} I. L. Bogolubsky and V. G. Makhankov, JETP Lett.
24 (1976) 12 [Pis'ma Zh. Eksp. Teor. Fiz. 24 (1976) 15].

\bibitem{gleiser} M. Gleiser, Phys. Rev. D {49}, 2978 (1994).

\bibitem{copeland} E. J. Copeland, M. Gleiser and H.-R. M\"uller, 
Phys. Rev. D \textbf{52}, 1920 (1995).

\bibitem{fodor-honda} G. Fodor et al, Phys. Rev. D \textbf{74}, 124003 (2006), E. P. Honda and M. W. Choptuik, Phys. Rev. D \textbf{65}, 084037 (2002).

\bibitem{graham_cos} N. Graham and N. Stamatopoulos, Phys. Lett. B
{\bf 639}, 541 (2006).

\bibitem{farhi_cos}
E. Farhi, N. Graham, A. Guth, N. Iqbal, R. Rosales and
N. Stamatopoulos, Phys. Rev. D {\bf 77}, 085019 (2008).

\bibitem{amin1d} M. A. Amin, arXiv:1006.3075.

\bibitem{3d_osc_temperature} M. Gleiser, N. Graham, N. Stamatopoulos, Phys. Rev. D \textbf{82}, 043517, (2010).

\bibitem{amin3d} M. A. Amin, R. Easther and H. Finkel, JCAP, 1012:001 (2010).

\bibitem{flat-top} M. A. Amin and D. Shirokoff, Phys. Rev. D \textbf{81}, 085045, (2010).

\bibitem{gleiser-thor} M. Gleiser and J. Thorarinson, Phys. Rev. D {\bf 76}, 041701(R) (2007); ibid., D {\bf 79}, 025016 (2009).

\bibitem{osc-gauge} N. Graham, Phys. Rev. Lett. {\bf 98}, 101801 (2007),
[Erratum-ibid. {\bf 98}, 189904 (2007)]; N. Graham, Phys. Rev. D {\bf
76} (2007) 085017.

\bibitem{WMAP} E. Komatsu \emph{et al}. [WMAP Collaboration] Astrophys. J. Suppl. \textbf{192}, 18 (2010).

\bibitem{sornborger} M. Gleiser and A. Sornborger, Phys. Rev. E\textbf{62}, 1368 (2000)

\bibitem{bellido-spectrum} J. Garcia-Bellido, D. Wands, Phys. Rev. D \textbf{54}, 7181 (1996). 

\bibitem{simulation} \href{http://www.youtube.com/watch?v=SWGgSM0Pacs}
{http://www.youtube.com/watch?v=SWGgSM0Pacs}
(Web link on electronic version).

\bibitem{gpu}N. Stamatopoulos (in preparation).

\bibitem{landau-lifshitz} L. D. Landau and E. M. Lifshitz, \emph{Mechanics}, Volume 1, 3rd Edition, Pergamon Press, (1976).

\bibitem{gleiser-howell} M. Gleiser and R. Howell, Phys. Rev. E {\bf 68}, 065203(RC)
(2003).

\end{thebibliography}
\end{document}